# Artificially built Kondo chains with organic radicals on metallic surfaces: new model system of heavy fermion quantum criticality

En Li[1], Bimla Danu[2], Yufeng Liu[3], Huilin Xie[4], Jacky Wing Yip Lam[4], Ben Zhong Tang[4,5], Shiyong Wang[3,6,7*], Fakher F. Assaad[2*], Nian Lin[1*]

[1]Department of Physics, The Hong Kong University of Science and Technology, Hong Kong, China

[2]Institut für Theoretische Physik und Astrophysik, Universität Würzburg, 97074 Würzburg, Germany

[3]Key Laboratory of Artificial Structures and Quantum Control (Ministry of Education), School of Physics and Astronomy, Shanghai Jiao Tong University, Shanghai 200240, China

[4]Department of Chemistry, Hong Kong Branch of Chinese National Engineering Research Center for Tissue Restoration and Reconstruction, The Hong Kong University of Science and Technology, Clear Water Bay, Hong Kong, China

[5]School of Science and Engineering, Shenzhen Institute of Aggregate Science and Technology, The Chinese University of Hong Kong, Shenzhen (CUHK-Shenzhen), Guangdong 518172, China

[6]Tsung-Dao Lee Institute, Shanghai Jiao Tong University, Shanghai, 201210, China

[7]Hefei National Laboratory, Hefei 230088, China

*Correspondence to: Nian Lin phnlin@ust.hk

## Abstract

**Heavy fermion quantum criticality is an extremely rich domain of research which represents a framework to understand strange metals as a consequence of a Kondo breakdown transition. Here we provide an experimental realization of such systems in terms of organic radicals on a metallic surface. The ground state of organic radicals is a Kramer's doublet that can be modeled by a spin ½ degree of**

**freedom. Using on-surface synthesis and scanning tunneling microscopy (STM) tip manipulation, one can controllably engineer and characterize chains of organic radicals on a Au(111) surface. The spatial-resolved differential conductance reveals site-dependent low-energy excitations, which support the picture of emergent many-body Kondo physics. Using quantum Monte Carlo simulations, we show that a Kondo lattice model of spin chains on a metallic surface reproduces accurately the experimental results. This allows us to interpret the experimental results in terms of a heavy fermion metal, below the coherence temperature. We foresee that the tunability of these systems will pave the way to realize quantum simulators of heavy fermion criticality.**

Quantum critical points (QCP) correspond to phase transitions where quantum fluctuations are dominant: the tuning parameter can be for instance pressure or magnetic field[1,2]. Although points in a phase diagram, they span critical fans in temperature that are controlled by the fluctuations of the QCP down to arbitrarily low temperatures[3]. Notable example are Kondo breakdown transitions and concomitant strange metals[1,4,5]. These heavy-fermion materials are characterized by local moments interacting with itinerant electrons, where the aforementioned tuning parameters tune the competition between the Ruderman-Kittle-Kasuya-Yosida (RKKY)[6] interaction and Kondo effect[7]. It is highly desirable to artificially build tunable systems that show the phenomenology of heavy fermion quantum criticality. It has recently been argued that this can be observed by considering one dimensional arrays of magnetic moments on metallic or semi-metallic surfaces[8-10]. The aim of this article is to propose a new experimental setup that can potentially pave the way to investigating heavy-fermion quantum criticality.

Experimentally, single-impurity Kondo physics, as well as finite-size spin lattices[6,11,12], have been studied by constructing arrays of magnetic atoms on a surface. Carbon-based magnetism, which arises from the presence of unpaired π-electrons in organic systems, exhibits remarkable properties including long spin coherence times and distances[13-16], large magnetic exchange couplings[14,15,17], and magnetic stability even at room temperature[15,17]. With the advancement of on-surface synthesis[18,19], such organic open-shell nanostructures provide a widely engineerable platform to explore various collective quantum magnetic behaviors[20-23]. For example, Roman et al. showed that S=1 chains exhibit Haldane phase[20] and realized the S=1/2 alternating-exchange Heisenberg model[21]. Wang et al. showed gapped excitation in S=1/2 antiferromagnetic chains, and topologically protected end states in S = 1 antiferromagnetic chains[22].

In these works, spin-spin exchange interactions dominate the spin-substrate Kondo interactions. Herein, we report a study of a model system in which the spin-spin exchange interactions and the spin-substrate Kondo interaction are competing. We employ on-surface synthesis and tip manipulation to construct a series of chains of S=1/2 π radicals with different length on a Au(111) surface. By using scanning tunneling microscopy/spectroscopy (STM/STS) together with density functional theory (DFT) calculations and quantum Monte Carlo (QMC) simulation, we demonstrate that these systems are well described by the phenomenology of a heavy fermion metallic state in which each Kramers doublet of the π radicals is entangled with the conduction electrons to form a new bound state carrying the quantum numbers of the electron[24]. Our systems are characterized by a Kondo temperature of the order of 60 K such that at the experimental temperature of 5K, they are well below the coherence temperature of the heavy fermion metal. We propose that these systems are tunable and offer the possibility to trigger quantum phase transitions belonging to the realm of heavy fermion quantum criticality.

**On-surface synthesis of molecular spin-1/2 chains on Au(111)**

As illustrated in Figure 1a, depositing the precursors **1**, dibromine–1-Methyl-2-(1,2,2-triphenylethenyl)benzene, onto a Au(111) substrate held at 473K leads to the scission of C-Br bond and formation of the covalently linked chains via on-surface Ullmann coupling. Figures 1b and S1 show the STM images of on-surface synthesized oligomers. The bright protrusions can be assigned to the methyl groups, comparing with the STM topography of tetraphenylethylene macrocycles[25]. Both *trans* and *cis* bonding configurations are found, with long chains mostly containing a mixed *trans*/*cis* structure (Figure S1). Further thermal treatment at 583K leads to the cyclodehydrogenation of the as-synthesized polymers. Figure 1d shows a section in a chain where the four units in the middle are planarized, while the two side units carry methyl groups. Two of the planarized units exhibit six distinct lobes (product **2**), while such features are absent in the other two units. We attribute the featureless appearance to the radical sites saturated with two hydrogen atoms, as depicted in Figure 1c. This is because, in the process of cyclodehydrogenation, the dissociated hydrogen atoms migrate on the surface and can saturate the radical sites. The DFT optimized structures of the radical/hydrogen-saturated unit are shown in Figure S2. As per Lieb's theorem[26], the sublattice imbalance in **2** induces one unpaired π-electron, with a magnetic ground state of S=1/2 as further confirmed by the spin-polarized DFT calculations (Figure S2d). Through STM tip manipulation, we can controllably dissociate one hydrogen atom from an $sp^3$ carbon site, thereby converting a closed-shell unit to an open-shell radical (**2**). Figure 1d-f present a manipulation process that converts the two closed-shell units to radicals, forming a radical tetramer chain step by step. Using this strategy, we construct series of radical chains with different lengths.

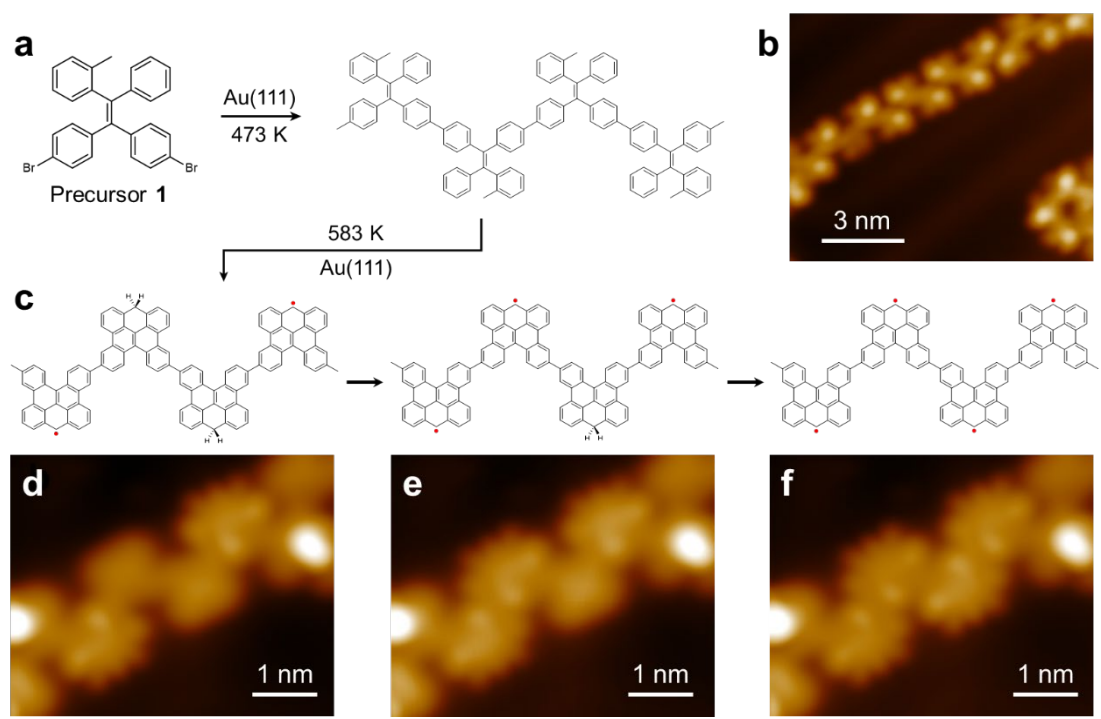

**Figure 1. On-surface synthesis and structural characterization of radical chains on Au(111). a**, Scheme of formation of the polymer chains using precursor **1**. **b**, High-resolution STM image (−1.7V,

100 pA) of the as- synthesized polymer chains with *trans*- (upper region) and *cis*- (bottom right region) configurations. **c**, Dissociation of the extra hydrogen atom from an $sp^3$ carbon site. Red dots represent the unpaired π electrons. **d-f**, High-resolution STM images (100 mV, 100 pA) showing the appearances of the same polymer chains during the atom manipulation.

**Radical monomer, dimer, and trimer**

The spin features of the radical chains are studied using STS. We start by characterizing the radical monomers. Figure 2a shows that in a polymer chain the central unit exhibits the radical features while the two side units are normal close-shell species. The differential conductance dI/dV spectra acquired at the central unit show a pronounced zero-bias peak, as shown in Figure 2b. The zero-bias peaks only exist on the central unit but not at the two side units. Figure 2c depicts the corresponding dI/dV map of Figure 2a acquired near zero bias, displaying a pattern with six distinct intensity lobes, which agrees well with the simulated constant-height STM image in Figure 2d and the calculated spin density distribution in Figure S2e. We attribute the zero-bias peak to an S=1/2 Kondo resonance. By fitting the spectra to a Fano function (Supplementary Section 3), we obtain a Kondo temperature $T_K = 64.9 \pm 3.2\ K$. By surveying 41 individual monomers, we found the $T_K$ ranges from ~ 34 to 118 K as shown in Figure S4. Statistically, the mean value is 57.3 K while the standard deviation (σ) is 10.5 K. The variation of $T_K$ is attributed to the fact that individual radicals may take different conformations depending on how they are linked with their neighbors in the chains, which results in different interaction strength of the spin to the substrate conduction electrons. It is worth mentioning that the $T_K$ here is about two to nine times larger than that in the reported organic radicals[14,22,27-29], indicating the radical interacts strongly with Au(111) surface electrons.

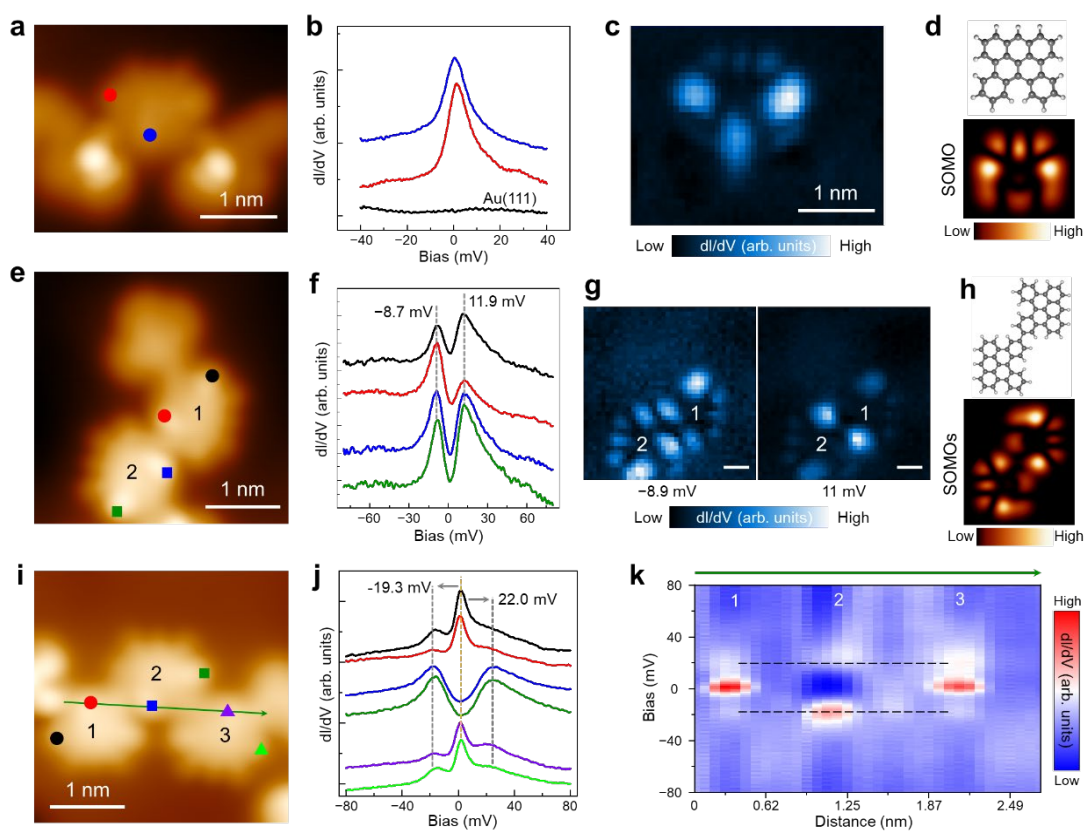

**Figure 2. Site-specific dI/dV spectra of a radical monomer, dimer, and trimer. a**, STM image (−100 mV, 1 nA) of a partially planarized oligomer containing one radical unit. **b**, dI/dV spectra taken on the locations marked in **a**. **c**, dI/dV map recorded at 1 mV in the same field of view as panel **a**. **d**, DFT optimized structure of a monomer (top) and the corresponding simulated constant-height STM image (bottom). **e**, STM image (−70 mV, 1 nA) of an oligomer containing two radical units. **f**, dI/dV spectra taken on the locations marked in **e**. **g**, dI/dV maps recorded at −8.9 mV and 11 mV in the same field of view as panel **e**, respectively. Scale bar: 0.5 nm. **h**, DFT optimized structure of a *trans*-dimer (top) and the corresponding simulated constant-height STM image (bottom). **i**, STM image (−80 mV, 100 pA) of a radical trimer. **j**, dI/dV spectra taken on the locations marked in **i**. **k**, dI/dV map taken along the green line in panel **i**, showing the absence of zero-voltage peak in the central unit of the trimer. The spectra in **b**, **f**, **j** are shifted vertically for clarity.

Figure 2e shows a radical dimer. As shown in Figure 2f, the site-specific dI/dV spectra, whose locations are labeled with the color dots, reveal that both radicals exhibit a double-peak spectral feature, with one peak at −8.7 mV and the other at 11.9 mV. The two peaks are not symmetric around zero bias, which excludes that the double-peak feature is an inelastic spin-flip excitation observed previously in diradicals[14,22,27,30]. The non-symmetric peaks indicate particle-hole asymmetry, which will be discussed below. Figure 2g shows that the dI/dV maps recorded at the energies of the two peaks display the same spatial patterns as the radical monomers (Figure 2c). The simulated constant-height STM image shown in Figure 2h matches the experimental data very well. Note that the observed double-peak feature is reproducible in dimers of different configurations (Figure S5). DFT calculations reveal an antiferromagnetic coupling in both *trans*- and *cis*-dimers, as shown in Figure S2f.

Figure 2i shows a radical trimer. Site-specific dI/dV spectra depicted in Figure 2j reveal a site-dependent behavior: the two side units exhibit a salient zero-bias peak and two satellite peaks at −17.9 mV and 23.4 mV. The center unit shows a U-shaped valley at zero bias and two peaks at slightly larger values than the two satellite peaks of the end units. The site-dependent spectra are better resolved in a line-mode dI/dV map shown in Figure 3k. One can see that the prominent zero-bias peak is located at the two end units (1&3) with two satellite peaks, while the center unit (2) features a valley at zero bias and two peaks above and below zero bias at slightly larger values than the satellite peaks of the end units as highlighted with the black dashed lines. Such site-dependent features are observed in all trimers regardless of their configurations. As shown in Figure S6, a trimer with a mixed *trans*/*cis* structure displays same site-dependent behavior as the *trans*-trimer (Figure 2i). Through step-by-step tip manipulation and characterization, the evolution of the spectra from a monomer to a trimer has further been confirmed, as shown in Figure S7.

**Longer radical chains**

We now discuss the spectral features of longer radical chains. Figures 3a and 3b show a tetramer and the corresponding site-specific dI/dV spectra. The zero-bias peak is absent in all four units. The two end units (1&4) display two pronounced peaks near zero bias at −3.3 mV and 5.2 mV. The two central units (2&3) exhibit two peaks at higher energies of −18.3 mV and 23.1 mV. The dI/dV maps shown in Figure 3c display spatial distribution of the states at −18.3 mV and −3.3 mV, respectively. The states at

−18.3 mV are distributed at the two middle units 2 and 3, while the states at −3.3 mV are distributed at two end units 1 and 4. In the radical pentamer (Figure 3d), site-specific dI/dV spectra (Figure 3e) show that the two end units (1&5) and the center unit (3) exhibit a zero-bias peak and two satellite peaks, while units 2 and 4 exhibit a valley at zero bias and two peaks at −21.9 mV and 25.7 mV for unit 2, −13.9 mV and 20.0 mV for unit 4. The radical hexamer (Figure 3f) does not exhibit zero-bias peaks at any unit as shown in Figure 3g. All units display peaks above and below zero bias. Overall, the even- and odd-numbered spin chains exhibit distinct spectral features: In the odd-numbered chains, the terminal units and the odd sites exhibit a zero-bias peak, while the even sites exhibit a double-peak feature; In the even-numbered chains, zero-bias peak is absent at all units, instead, all unit exhibit multiple peaks above and below the zero bias.

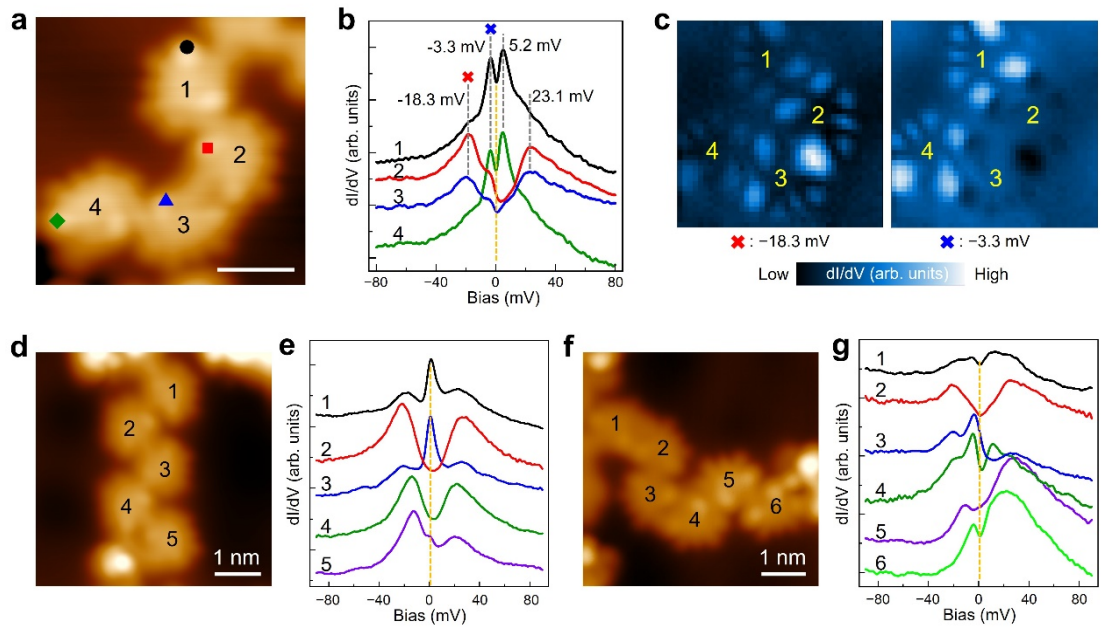

**Figure 3. Site-specific dI/dV spectra of radical tetramer, pentamer, and hexamer**. **a**, STM image (−80 mV, 1 nA) of a radical tetramer. Scale bar: 1 nm. **b**, dI/dV spectra taken on the locations marked in **a**. **c**, dI/dV maps recorded at −18.3 mV and −3.3 mV in the same field of view as panel **a**, respectively. **d**, STM image (−90 mV, 1 nA) of a radical pentamer. **e**, Corresponding dI/dV spectra taken on each unit in **d**. **f**, STM image (−30 mV, 30 pA) of a radical hexamer with a mixed *trans/cis* structure. **g**, Corresponding dI/dV spectra taken on each unit in **f**. The orange dashed line marks the dip/peak position of the spectra near the zero bias. All the spectra in **b**, **e**, **g** are shifted vertically for clarity.

**Theoretical modelling and quantum Monte Carlo simulations**

The length-dependent behavior observed in this work is different from the reported S=1/2 radical chains[22], which show spin excitation gaps for all the lengths and can be explained by the Heisenberg antiferromagnet S=1/2 model. Here we interpret the observed low-energy excitations in terms of Kondo physics, where the metallic surface is a relevant perturbation that has to be included in the modeling. An accurate modeling of the system should take into account the $p_z$ orbitals of the π radicals, the Coulomb repulsion as well as the detailed coupling to the substrate. This corresponds to a periodic Anderson model. Generically, such a modeling breaks particle-hole symmetry as a consequence of various effects, such as hybridization between the radical and the metallic surface, the lack of particle-hole symmetry of the metallic surface and disorders. Since considering particle-hole asymmetry is not central for the low energy modeling, we will neglect it. Importantly, this allows to avoid the negative sign problem in quantum Monte Carlo simulations[31]. For our low energy model, as illustrated in Figure 4a, we adopt a Kondo lattice Hamiltonian: the π radical is described by a Kramers doublet on which the spin operator $\hat{S}_l$ acts and the Au(111) surface is modeled by a fermi liquid of conduction electrons; $\hat{c}^\dagger_{i,s}$. This operator creates an electron in Wannier state centered around site $i$ and with z-component of spin s. The model Hamiltonian reads:

$$\hat{H} = -t \sum_{\langle i,j \rangle, s} \left( \hat{c}^\dagger_{i,s} \hat{c}_{j,s} + \text{H.c.} \right) + \frac{1}{2} \sum_{l=1,s,s'}^{L} J_l^k \hat{c}^\dagger_{l,s} \boldsymbol{\sigma}_{s,s'} \hat{c}_{l,s} \cdot \hat{\boldsymbol{S}}_l + \sum_{l=1}^{L-1} J_l^h \hat{\boldsymbol{S}}_l \cdot \hat{\boldsymbol{S}}_{l+\Delta l}$$

Here, t is the hopping parameter, $J_l^k$ the site-dependent antiferromagnetic Kondo coupling between the Kramer doublet and conduction electrons, $J_l^h$ the bond-dependent Heisenberg antiferromagnetic coupling, L the length of the Heisenberg chain, $\sigma$ a vector of Pauli spin matrices, $\langle i,j \rangle$ the nearest neighbors of a square lattice with lattice vectors $e_x, e_y$, and $l = n\Delta l$ with $\Delta l = 2e_y$. Throughout the calculation we set t=1, $J_l^k$=2, $k_BT$=0.1 and use a 20×20 lattice with periodic boundaries for the conduction electrons. Simulations were carried out with the ALF[32] implementation of the finite temperature auxiliary field quantum Monte Carlo algorithm[33,34] and ALF implementation of the Classic Maximum Entropy method[35] to carry out the analytical continuation. We have used an annealing procedure in which the spectral function of the higher temperature data acts as default model for the analytical continuation at the lower temperature. To compare with the experiments, we have computed the local spectral function, $A_\Psi(l,\omega)$, of the co-called composite fermion operator, $\hat{\Psi}^\dagger_{l,s} = \sum_{s'} \hat{c}^\dagger_{l,s'} \boldsymbol{\sigma}_{s',s} \cdot \hat{\boldsymbol{S}}_l$, that carries the quantum numbers of the electron. The hallmark of the Kondo effect is the emergence of a resonance in $A_\Psi(l,\omega)$, or new quasiparticle, below the Kondo temperature (Figure S8)[36,37].

As mentioned above, our model has built in particle-hole symmetry such that the spectra are symmetric around the Fermi level. Figure 4b-e shows the spectral function of L=2, 3, 4 and 5 spin chains, respectively. The Heisenberg antiferromagnetic coupling, $J^h_l$, are chosen to match the simulated line shapes as best as possible with the experimental dI/dV spectra. Upon inspection one observes remarkable agreement with the experimental data. The interpretation of the data stems from the Kondo effect for independent impurities where each spin degree of freedom is entangled with conduction electrons so as to generate the aforementioned composite fermion operator. In this framework, $J^h_l$, corresponds to a hopping matrix element of the composite fermions. At L = 2, this produces a bonding antibonding band and the observed gap. At L=3 the data can be understood in terms of three composite fermions on an open chain with hopping matrix element set by $J^h_l$. A calculation shows that the zero-mode wave function has a support centered on site 1 and 3 corresponding to the position at which we observe the peak at vanishing energies. At L=4, a similar picture holds. Since the ground state of four electrons hoping on an open four site chain is non-degenerate, we expect gaps in the spectra irrespective of the site. To reproduce the experimental spectra quantitatively, we take $J^h_l$ to be smaller on the edge bonds. This choice reduces the size of the gaps at the ends of the chain. The picture emerging from comparison between experiment and numerics is nothing but the large-N mean-field theory of heavy fermions[38] that has numerically be shown to be valid for the heavy fermion phase down to N=2[39]. Hence, we can conclude that the experiments correspond to the realization of a heavy fermion phase. Note that the different $J_l^h$ values may stem from the configuration variation of the polymer chains, specifically, the different dihedral angles between the neighboring radicals[40]. DFT calculations (Figure S3) show that the singlet-to-triplet gap varies from 7.4 meV to 15.8 meV when the dihedral angle changes from 50° to 0°, which correspond to the range of the antiferromagnetic coupling strength.

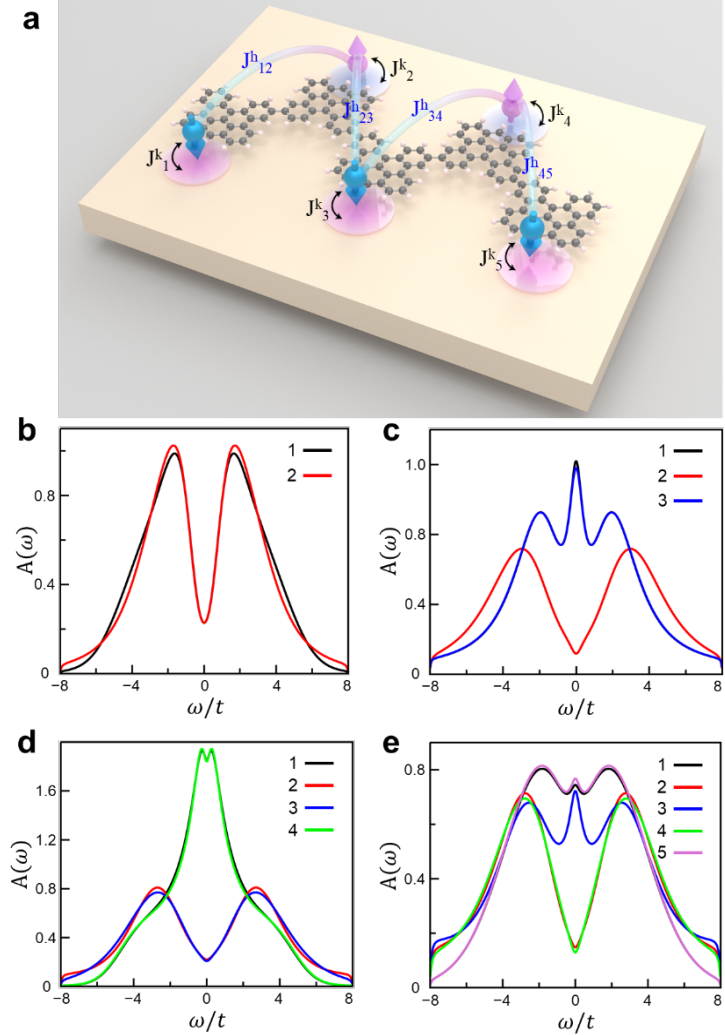

**Figure 4. The schematic theoretical model and spectral functions of the composite fermion operator at various sites of radical chains computed using the classical analytical continuation algorithm. a**, Schematic illustration of the theoretical model of experiment setup. The arrows represent the spins from the radicals, and colored regions represent the spins of conduction electrons coupled to the radical spins. $J^k$ is the Kondo coupling, whereas $J^h$ is the Heisenberg antiferromagnetic coupling between radical spins. **b**, Dimer: L=2, $J^h_{12}$=1.5. **c**, Trimer: L=3, $J^h_{12}$=$J^h_{23}$=1.5. **d**, Tetramer: L=4, $J^h_{12}$=$J^h_{34}$=1, $J^h_{23}$=2. **e**, Pentamer: L=5, $J^h_{12}$=$J^h_{23}$=$J^h_{34}$=$J^h_{45}$=1.8. All spectra are computed at $J_k/t = 2$ and

$\beta t = 10$.

In summary, we have demonstrated that one can artificially build Kondo lattice systems; STM tip manipulation can selectively and controllably dissociate one hydrogen atom from an *sp*³ carbon site, thereby converting a closed-shell unit to an open-shell radical. As it stands, experiments point to the realization of heavy fermion phenomenology in which each Kramers pair forms an entangled state with conduction electrons leading to a new excitation carrying the quantum numbers of the electron. Key questions for accessing heavy fermion quantum criticality are the size of the system and tunability. Numerical simulations show that building small rings of the order of ten to twenty Kramers pairs already suffices to pick up aspects of quantum criticality. Experimentally, one can design precursor molecules that afford longer/shorter radical-to-radical separation or larger/smaller dihedral angle, both can change the antiferromagnetic coupling strength[17,40]. On the other hand, one can reduce the Kondo strength by designing precursor molecules with bulky side groups to alter the molecule-to-substrate distance. We conjecture that these systems will be a promising avenue to investigate heavy fermion criticality. Furthermore, it would be worthwhile to investigate how disorder affects quantum critical behavior in these systems theoretically due to the inherent disorder effects associated with the Kondo scale in these systems.

## Methods

**On-surface synthesis and STM/STS measurements.**

All the experiments were carried out in an ultrahigh vacuum system (base pressure $<3.0 \times 10^{-10}$ mbar) equipped with a CreaTec low-temperature STM. The Au(111) single-crystal surface was prepared through cycles of Ar+ sputtering (0.8 keV) and subsequent annealing at 750 K. Powder samples of precursor **1** were sublimed from a Knudsen cell at 413 K, while the substrate was kept at 473 K during molecule deposition. After further annealing at 583 K, the sample was subsequently transferred to a cryogenic scanner for characterization. STM/STS measurements were performed at 5.3 K using a chemically etched tungsten tip. Differential conductance (dI/dV) spectra were collected by using the standard lock-in technique with a voltage modulation of 0.6-1 mV and frequency of 727.3 Hz. STM images and dI/dV maps were recorded in constant-current modes, while dI/dV spectra were recorded in constant-height mode. dI/dV spectra on bare Au(111) substrate were used as an STS reference for tip calibration. To conduct the atom manipulation, we first positioned the STM tip above the $sp^3$ carbon site. The bias voltage was then gradually ramped to 2.5-3.0 V until an abrupt change in the tunneling current was observed, indicating the dissociation of one hydrogen atom.

**DFT calculations**

Spin-polarized DFT calculations were conducted using the Gaussian 16 package. The PBE0-D3 (BJ) functional was employed to illustrate the electronic structure of gas-phase molecules[41], with the def2-SVP basis set utilized for geometry optimization. This was further extended to a def2-TZVP basis set for the single-point energy calculation[42]. Images of the structures and spin density distributions were plotted by IQmol molecular viewer.

**Quantum Monte Carlo**

We have used the ALF-implementation[32] of the finite temperature auxiliary field quantum Monte Carlo method[34,33,43] to compute thermodynamical and dynamical properties of the model described above. The formulation of the algorithm follows the work of Refs.[44,45] and the details of the implementation can be found in the ALF-documentation[32]. The electron tunneling from the STM tip to the metallic surface can occur directly or indirectly through the π radical. Considering that experimentally a peak is observed, the latter channel dominates. This process is described by the composite fermion spectral function described in the main text. For our simulations we have used an imaginary time step $\Delta\tau t = 0.1$. The spectral functions are obtained from the annealing process described in the main text carried out in the inverse temperature range $\beta t = 1$ to $\beta t = 10$

## Acknowledgements

E.L. and N.L. thank the Hong Kong RGC (project 16301423). B.D. and A.F.F. gratefully acknowledge the scientific support and HPC resources provided by the Erlangen National High Performance Computing Center (NHR@FAU) of the Friedrich-Alexander-Universität Erlangen-Nürnberg (FAU) under NHR project 80069 provided by federal and Bavarian state authorities. NHR@FAU hardware is partially funded by the German Research Foundation (DFG) through grant 440719683. They also gratefully acknowledge the Gauss Centre for Supercomputing e.V. (www.gauss-centre.eu) for funding this project by providing computing time through the John von Neumann Institute for Computing (NIC) on the GCS Supercomputer JUWELS[46] at Jülich Supercomputing Centre (JSC). A.F.F acknowledges financial support from the German Research Foundation (DFG) under the grant AS 120/16-1 (Project number 493886309) that is part of the collaborative research project SFB Q-M&S funded by the Austrian Science Fund (FWF) F 86. BD acknowledges financial support from the German Research Foundation (DFG) under the grant DA 2805/2 (Project number 528834426). S.W. thank the Ministry of Science and Technology of China (2020YFA0309000), NSFC (Grants No. No. 22325203), and the financial support from Innovation program for Quantum Science and Technology (Grant No. 2021ZD0302500).

## Author contributions

N.L. and F.F.A. supervised the project; E.L. performed the on-surface synthesis and STM/STS experiments; B.D. and F.F.A. carried out the QMC simulations; H.X. synthesized the precursor molecules under the supervision of J.W.Y.L. and B.Z.T.; Y.L. and S.W. performed DFT calculations; the manuscript was written by E.L., F.F.A., and N.L. with contributions from all co-authors. All authors contributed to the scientific discussion.